\documentclass[amsmath,amssymb,aps,twocolumn,jcp,floatfix]{revtex4}

\usepackage{mathrsfs}
\usepackage{amsmath}
\usepackage{amssymb}
\usepackage{color}
\usepackage{graphicx}	
\usepackage{bm}
\usepackage{ulem} 
\usepackage{titlesec}
\titleformat{\paragraph}[runin]
{\bfseries\scshape}{\theparagraph}{1em}{}
\usepackage{braket}
\usepackage{multirow}

\usepackage[printwatermark]{xwatermark}
\usepackage{xcolor}

\newcommand{\be}{\begin{equation}}
\newcommand{\ee}{\end{equation}}
\newcommand{\bef}{\begin{figure}}
\newcommand{\eef}{\end{figure}}
\newcommand{\bea}{\begin{eqnarray}}
\newcommand{\eea}{\end{eqnarray}}

\begin{document}
\title{Sub-Arrhenius diffusion in a binary colloidal system
}
\author{Mahammad Mustakim and A. V. Anil Kumar}
\thanks{Corresponding author: \texttt{anil@niser.ac.in}.}
\affiliation{School of Physical Sciences, National Institute of Science Education and Research, HBNI, Bhubaneswar-752050, India}

\date{\today}
\begin{abstract}
 The dynamics of binary colloidal mixture subjected to an external potential barrier is investigated using molecular
 dynamics simulations. The depletion interactions between the potential barrier and larger components of the mixture 
 alters the dynamics of the system significantly. The larger particles undergo sub-Arrhenius diffusion while smaller
 particles obey normal Arrhenius diffusion. These results show that quantum phenomena such as tunneling is not required to have sub-Arrhenius
 diffusion, in contrast to the general agreement in the literature. The depletion interactions between the external
 potential barrier and larger component increases with decreasing temperature which makes the effective activation energy for barrier crossing temperature 
 dependent leading to sub-Arrhenius diffusion. 

\end{abstract}
\maketitle

\section{Introduction}

Understanding thermally activated escape over a barrier is at the heart of many important phenomena
such as transport processes in living cells, polymeric solutions, colloidal systems, conformational diffusion 
in proteins, chemical reactions etc. to name a few. Since the seminal work of Hendrik A. Kramers in 1940~\cite{Kramers}, this problem attracted 
lot of attention in different fields of science\cite{Hangii}. Kramers treated this problem as the escape of a Brownian 
particle from a metastable state and obtained the escape rate over a potential barrier subjected to Gaussian
white noise. Temperature dependence of this escape rate or diffusion coefficient is given by Van't Hoff - Arrhenius - Kramers
formula\cite{vanthoff1884,Arrhenius1889},
\begin{equation}
 D(T) = D_0 e^{(-E_a/k_BT)}
\end{equation}

\noindent This equation was first proposed by Van't Hoff and Arrhenius and later derived by Kramers by solving the Fokker-Plank
equation for Brownian motion in phase space in the presence of a nonlinear potential function. The Van't Hoff-Arrhenius
law is found to be robust in the description of diffusion coefficient or rate constant in many processes such as
particle diffusion in solids and liquids\cite{huger,purdue}, diffusion in microporous materials\cite{karger,garcia}, elementary chemical 
reactions\cite{zellner,kohen}, enzymatic catalysis\cite{swanwick,gul}, electrical conductivity in ionic liquids and super ionic conductors\cite{kamaya,wagner} etc. However, 
deviations from Van't Hoff - Arrhenius equation are observed in many systems especially at low temperatures. In general, these deviations
are attributed to a temperature-dependent activation energy in contrast to the energy barrier independent of
temperature in the Arrhenius picture. These deviations are 
generally classified into two: super-Arrhenius and sub-Arrhenius diffusion depending  the convex or concave nature of
the log($D$) versus $1/T$ curves. They correspond to an increase or decrease in the activation energy as temperature 
decreases.

 Recently a simple formalism was proposed\cite{aquilanti1,aquilanti2,silva1,silva2,nishiyama1}, inspired by Tsallis nonextensive statistical mechanics\cite{tsallis} to describe the 
deviations from Arrhenius behaviour in diffusivities or rate constants in terms of a single parameter. Exploiting Euler's
expansion for exponential function as the limit of a succession, this formalism proposes a deformed-Arrhenius
equation for the temperature dependence of diffusion coefficient,

\begin{equation}
D(T) = A\Big[1-d\frac{E_0}{k_BT}\Big]^{1/d}
\end{equation}

This formula respresents both Arrhenius and non-Arrhenius behaviour of diffusion. Here $E_0$ is the height of
the potential barrier and $d$ is known 
as the deformation parameter, the sign of which will determine the nature of deviations from Arrhenius behaviour. 
Then the activation energy for diffusion is defined as
\begin{eqnarray}
 E_a = -\frac{d ln D}{d(1/k_BT)} &=& E_0\Big(1-\frac{E_0d}{k_BT}\Big)^{-1} \nonumber \\
 &\cong& E_0 + d\frac{E_0^2}{k_BT}\, \,  (\textrm{for small} \, \, d)
\end{eqnarray}

For positive values of $d$, super-Arrhenius behaviour is observed and sub-Arrhenius behaviour is observed 
for negative values of $d$. For $d$ = 0 equations (2) and (3) tend to the form of Arrhenius equation. Super-Arrhenius behaviour 
of diffusion is mainly found in systems where collective or cooperative dynamics is predominant such as dynamics
of supercooled liquids\cite{Stirnemann, desouza}, diffusion through membranes\cite{rotella,nishiyama}, 
macroscopic sliding of bacteria\cite{chen} etc. Meanwhile, sub-Arrhenius behaviour
is mainly seen in chemical reactions\cite{aquilanti, liang, braun} and quantum tunneling was proposed to be responsible for such behaviour. In fact, 
Bell model for incorporate tunneling in chemical kinetics was extended to correlate the deformation parameter $d$ 
with the parameters of the energy barrier\cite{bell1,bell2}. Therefore in the
literature, there is 
a consensus that super-Arrhenius behaviour occurs in classical systems, while sub-Arrhenius behaviour occurs at 
processes where quantum tunneling plays a significant role.  To the best of our knowledge, there has been no investigations
reported in the literature of a classical system exhibiting sub-Arrhenius diffusion. However, sub-Arrhenius behaviour has been observed in the 
sedimentation of weekly-aggregated colloidal gels\cite{buscall}. In this article, we report
a purely classical system of binary colloidal mixtures, subjected to an external potential energy barrier, 
undergoing sub-Arrhenius diffusion. We show that one of the components 
in the binary mixture undergoes sub-Arrhenius diffusion, while the other component exhibits normal Arrhenius diffusion.
This contrasting behaviour of different components in the mixture can be attributed to the attractive depletion interaction between 
the potential energy barrier and the larger component in the mixture.

\section{Model and simulation details}

We have carried out canonical ensemble molecular  dynamics simulations of a binary mixture of colloidal particles, 
 the two components of the mixture differ in their sizes, subjected to an external potential barrier. 
 The interaction potentail,
 $V_{ab}(r_{ij})$, between these particles is soft-sphere repulsion, given by
 
\begin{equation}
 V_{ab}(r_{ij}) = \epsilon_{ab} \Big(\frac{\sigma_{ab}}{r_{ij}}\Big)^{12}
\end{equation}
 
\noindent where $r_{ij}$ is the distance between two particles $i$ and $j$ and $a$, $b$ = $l$, $s$, where $l$ stand for the the larger particles and $s$ for the smaller particles.
In our simulations, $\sigma_{ss}$ = 1.0, $\sigma_{ll}$ = 2.0, $\epsilon_{ss}$ = 1.0 and $\epsilon_{ll}$ = 4.0 all being expressed
in reduced units. The parameters for interaction between the unlike species is determined by the Lorenz-Berthelot mixing rules; i.e., $\sigma_{sl} = (\sigma_{ss} + \sigma_{ll})/2.0$
and $\epsilon_{sl} = \sqrt{\epsilon_{ss}\epsilon_{ll}}$. The system is subjected to an external potential which is in the
form of gaussian barrier at the center of the simulation
box along the z-axis\cite{anil,anil1}
\begin{equation}
V_{ext}(z) = \epsilon_{ext} \,\, e^{-\Big(\frac{z-z_0}{w}\Big)^2}
\end{equation}
We chose the width of the external potential to be $w$ = 3.0 and the height of the barrier to be 
$\epsilon_{ext}$ = 3.0. Recent experimental advance make it possible to realise such 
external potentials in colloidal systems\cite{thorn}. As in ref.\cite{anil1}, we use an equivolume mixture 
with a total volume fraction of $\phi$ = 0.20.
We kept the masses of both the species same since we are interested to see only the effect of depletion interactions in the dynamics.
The dynamics of this system has been investigated using canonical ensemble 
molecular dynamics simulations. The equations of mation are solved simultaneously using a fifth order Gear predictor-corrector 
method\cite{allen}, applying periodic boundary conditions along all the three directions. We have repeated the simulations 
for three different simulation boxlengths, i.e., $L$ = 13.0, 15.0 and 17.0 to study the finite size effects on 
the dynamics, keeping the total volume fraction to be 0.20. 
Because of the periodic boundary conditions, different boxlengths essentially mean different periodicity
for the external potential barrier. Each simulation runs for a total of 5 $\times$ 10$^6$ timesteps with each timestep being $dt$ = 0.001
in reduced units, in which first 1 $\times$ 10$^6$ 
are not used in calculating the equilibirium and dynamic properties. The simulations are repeated three times for each 
set of parameters and the dynamical properties are averaged over.

 This model system has been investigated before to establish that the presence of smaller particles invoke depletion interactions
between the larger particles as well as between the larger particles and the potential barrier\cite{anil,anil1}. It has been shown that this 
attractive depletion interaction significantly alter the structural and dynamical behaviour of the particles. 
At low enough temperatures, the smaller particles 
get localised between the potential barriers (multiple barriers occur because of the periodic boundary conditions) 
and undergo a slowing down of dynamics, while the larger particles diffuse normally without acknowledging the 
presence of the repulsive barrier\cite{anil1}. It may also be noted that the single component systems
undergo a slowing down in their dynamics in both the case of larger and smaller particles(see additional information).
This striking dynamical behaviour has been attributed to the changes in 
the effective potential barrier due to the depletion interaction
between the larger particles and the potential barrier.
We attempt to further 
understand the nature of depletion interaction and its effect on the dynamics of the system, especially at low temperatures where changes in dynamical properties are 
significant. 

\section{Results and discussion}

\begin{center}
\begin{figure}
\includegraphics[width=6cm, angle=-0]{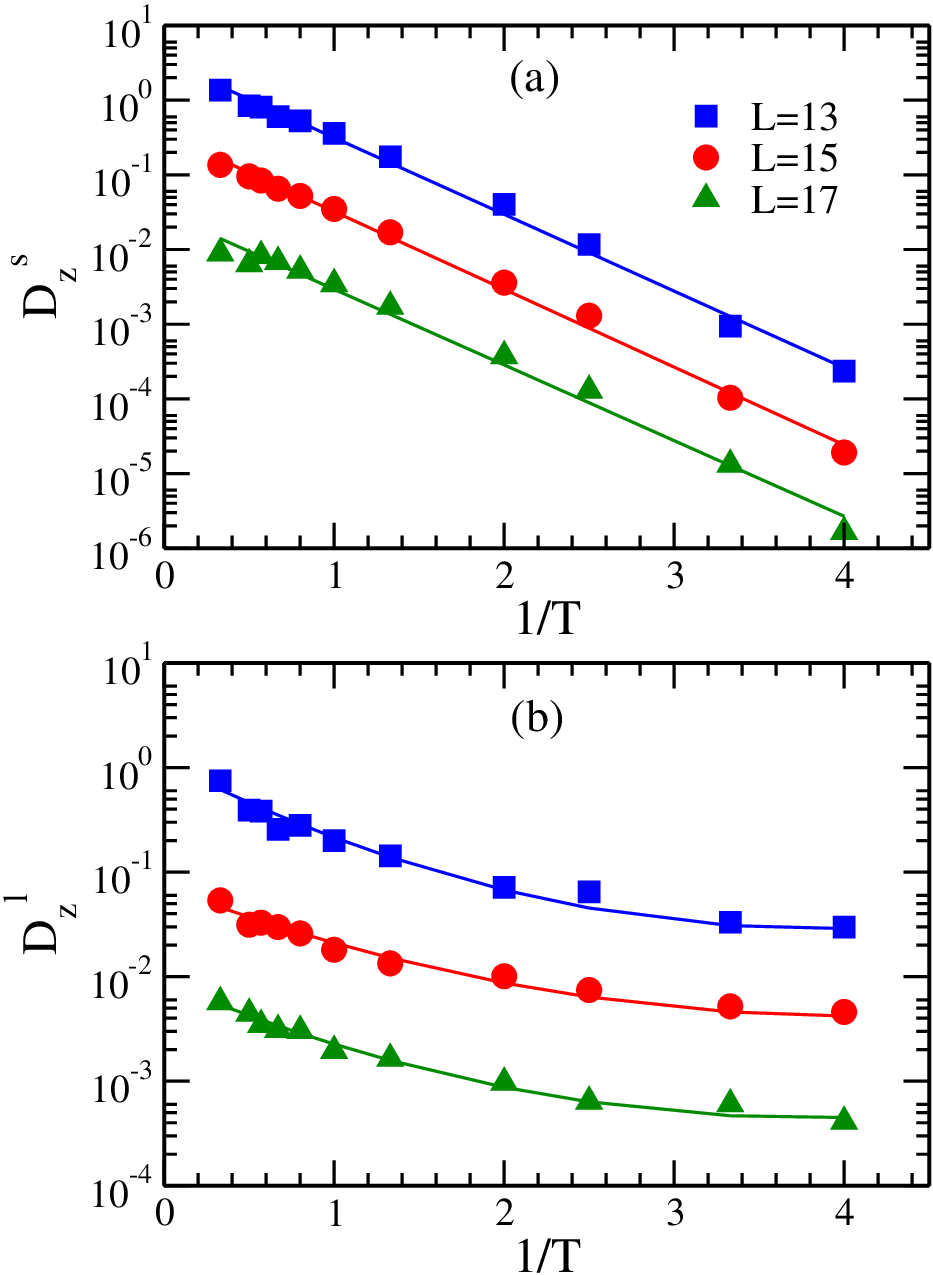}
\caption{Arrhenius plot for (a) small particles and (b) large particles. Small particles follow Arrhenius law, while 
larger particles follow sub-Arrhenius behaviour. The points are from simulation data and the lines are fits to Arrhenius
or $d$-Arrhenius law. The data points for $L$ = 15 and $L$ = 17 shifted by 1 and 2 orders of magnitude.}
\label{fig1}
\end{figure}
\end{center}

 As we decrease the temperature of this system, the self-diffusion coefficient of different species shows markedly different 
 behaviour. While the diffusion coefficient ($D_z^s$) of smallar particles along the $z$-direction
 decreases rapidly as temperature decreases, the larger particles
 diffusivity($D_z^l$) decrease rather very slowly (see figure 3(a) in the additional information).
This slow change in the self-diffusion 
coefficient of larger particles suggests that the activation energy for their dynamics is temperature-dependent 
and  the diffusion can be non-Arrhenius. In order to verify this, we 
calculated the activation energy of diffusion for both the species of particles. 
 We have also carried out two additional simulations - 
one with larger particles only and another with smaller particles only. 
From these simulations we calculated the 
activation energy for each of the components in the absence of depletion interactions. This will enable us to calculate the change in activation energy due to the 
depletion interaction between the potential barrier and the larger particles. The diffusion is found to be 
Arrhenius for both single component systems. The activation energy for smaller particles is found to be 2.1066 and that for larger 
particles is 1.8984 (See additional information).  Figure 1 shows 
 the Arrhenius plot for both the species of particles for different values of simulation boxlengths (different values of simulation boxlength means different
 periodicity for the external potential barrier). It is evident from the figure that the smaller particles
 obey Arrhenius behaviour for all boxlengths. The activation energy is found to be around 2.4 for all 
 the boxlengths, greater than that found in the case of smaller particles alone. This agrees with the suggestion made in 
 ref.\cite{anil1} that the depletion interaction effectively increases the barrier for smaller particles as they are 
 driven away from the barrier. The Arrhenius plot for larger particles is significant that the plot is no longer linear
 and it deviates from Arrhenius behavior. In fact we can see that they show a sub-Arrhenius diffusion at all periodicities of the external potential.
 As mentioned before, sub-Arrhenius behavior is observed so far only in systems where quantum tunneling is observed
 and believed to be a quantum phenomena. To the best of our knowledge, this is the first classical system to
 be reported to have a sub-Arrhenius behavior. Moreover, the two components in the binary mixture follows different 
 temperature dependence in their diffusional behaviour.

 \begin{center}
\begin{figure}
\includegraphics[width=6cm, angle=-0]{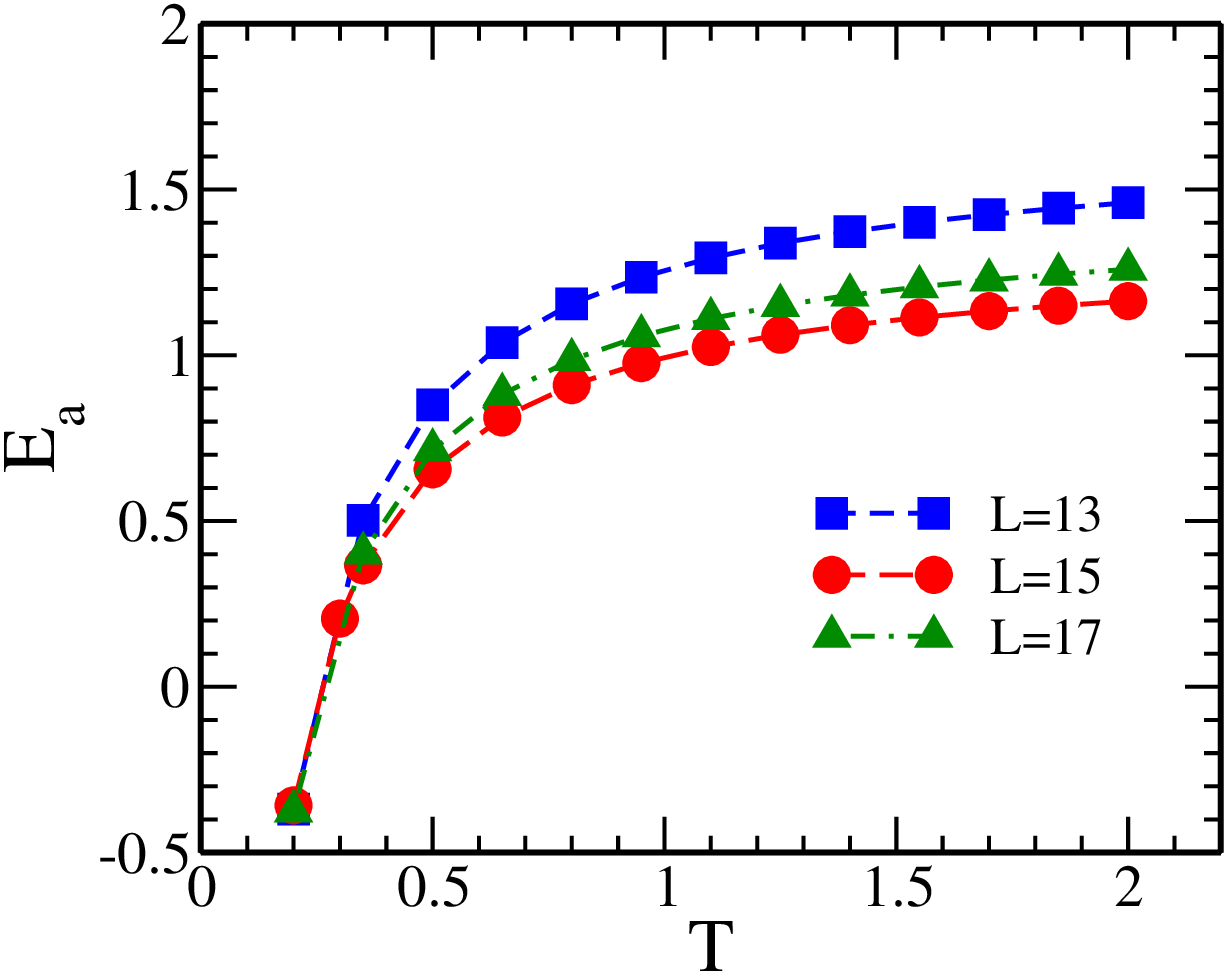}
\caption{Activation energy versus temperature for larger particles.}
\label{fig2}
\end{figure}
\end{center}

The Arrhenius plot of diffusion of larger particles is found to obey the $d$-Arrhenius equation as revealed by the nonlinear 
fits to the diffusivity data shown in Figure 1. The temperature dependent activation energy can be calculated 
from the $d$-parameter using equation 3. This has been plotted at Figure 2 for different boxlengths. The activation 
energy is less than the activation energy for diffusion when only large particles are present at all temperatures. 
This is in agreement with the suggestion that the attractive depletion interaction between
the potential barrier and larger particles effectively lowers the barrier for the larger particles to cross. 
The reduction in the activation energy is larger as the temperature is lowered. This essentially means the depletion 
interaction is temperature dependent and become more and more prominent as temperature decreases. This also 
explains why the diffusion coefficient remains same or changes very slowly as temperature is lowered. The diffusivity 
depends on two factors here; namely activation energy and temperature or more precisely the ratio of activation energy to temperature.
In case of smaller particles the activation energy remains constant; so the ratio of activation energy to temperature becomes larger as temperature decreases. This reduces the 
diffusivity rapidly. In case of larger particles, the activation energy decreases as temperature lowers so that the ratio changes very little at different temperature. 
This in turn ensures that
the diffusivity of larger particles remains more or less same irrespective of the changes in temperature. 

\begin{figure}
\includegraphics[width=6cm, angle=-0]{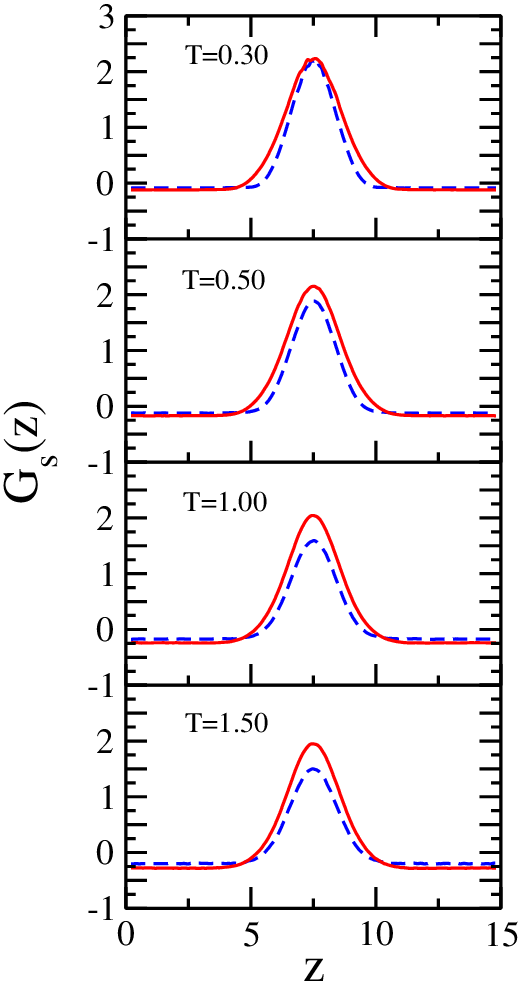}
\caption{Free energy of interaction for the small particles along the $z$ - direction when the system has only small particles(dashed lines) and when both large and small particles
present (solid lines).}
\label{fig3}
\end{figure}

In order to gain further insight into the mechanism of this interesting behaviour in the dynamics of the 
 particles in the binary mixture, we calculated the free energy of interaction along the z-direction. This 
 is obtained from the 
 density profile of each of the components along the $z$-direction, using\cite{nygard}
 \begin{equation}
 G(z) = -k_BT ln \, n_i(z)
 \end{equation}
\noindent where $n_i(z)$ is the density profile for each of the species along the z-direction normalised
with respect to the bulk density. 
 Figure 3 shows the free energy of interaction for smaller particles in two different systems; namely only smaller 
particles with the external potential barrier(dashed line) and the binary mixture with external potential barrier 
(solid lines) at four different temperatures.  When only small particles are present, there is no depletion 
interaction and barrier does not get modified. 
However, the binary mixture invokes depletion interaction between
the barrier and the larger particles.  So the density of larger particles near the barrier increases and because 
of this crowding, the smaller particles
will move away further from the barrier. This increases the effective free energy for smaller particles, which 
is evident in Figure 3.
However, since the volume fraction is small, this decline in the density of smaller particles
from the region of external potential barrier does not depend on temperature and 
hence the free energy of interaction remains unchanged with respect to temperature. So the effective potential 
barrier is temperature independent and the diffusion process remain Arrhenius. However at higher volume fractions, 
the effective barrier become temperature dependent and the diffusion becomes super-Arrhenius\cite{mustakim}. Figure 
4 shows the free energy profile of larger
particles for two different systems : (1) only larger particles with external potential barrier and (2) the binary 
mixture with external potential barrier at four different temperatures. There are few points to be noted here. 
Firstly, for larger particles in the binary mixture, there is a minimum for free energy of interaction at
the spatial position of the external potential barrier rather than  a maximum, which is observed in the
case of single component system or in the case of smaller particles in the binary mixture. Here the free energy 
barrier particles have to face in their dynamics is the difference in free energy between the minimum and bulk(free energy
value away from the external potential barrier). 
This barrier height is much smaller compared to the barrier height particles have to cross in single component system. 
Secondly, for the single component system the barrier height does not change significantly  with respect to 
temperature while the 
barrier height decreases with decreasing temperature in case of binary mixture. These observations are consistent 
with the conclusions made based on the activation energy calculations outlined above. The fact that the free energy 
barrier is temperature dependent for the binary mixture and that the barrier height decreases with decreasing 
temperature provides an explanation for the sub-Arrhenius diffusion observed in the case of larger particles. 
\begin{figure}
\includegraphics[width=6cm, angle=-0]{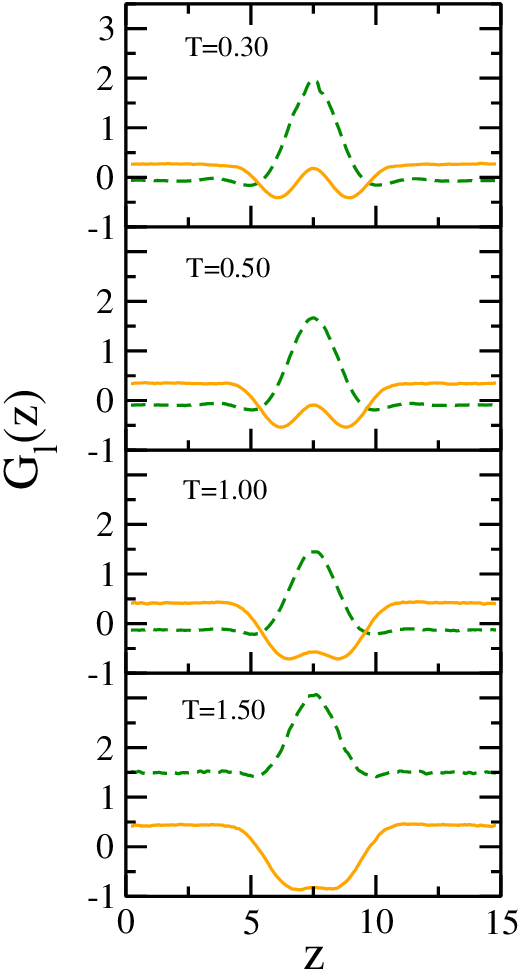}
\caption{Free energy of interaction for the larger particles along the $z$ - direction when the system has only large particles(dashed lines) and when both large and small particles
present (solid lines).}
\label{fig4}
\end{figure}

 From the density profile, we can also calculate the mean force as\cite{nygard}
 \begin{equation}
 F(z) = k_BT \frac{d( ln \, n_i(z))}{dz}
 \end{equation}
\noindent along the direction of the potential barrier. Here $F(z) = F_+(z) - F_-(z)$, where $F_+(z)$ is the mean force
of interaction along positive $z$ direction and $F_-(z)$ is the mean force along negative $z$ direction.
These are plotted in Figure 5 and 6 for smaller particles 
and larger particles respectively.  From figure 5 it is clear that the force on smaller particles near the 
potential barrier is repulsive (negative on the left side of the barrier and positive on the right side) both 
in the single component system and in the binary mixture.  
Also it is clear that the magnitude of the force does not change significantly with
respect to temperature and hence the dynamics remains Arrhenius. However, the effective force in the larger 
particles shows a very contrasting behaviour. The force is mainly attractive towards the barrier (positive on 
the left of the barrier and negative on the right). And the magnitude of the force changes as the temperature 
changes. This confirms the sub-Arrhenius behaviour obeserved in the diffusivity of larger particles. It should 
be noted that the fluctuation in 
the mean force in the region of potential barrier at low temperatures arises due to the splitting of the peak in 
density profile at lower temperatures.

\begin{figure}
\includegraphics[width=6cm, angle=-0]{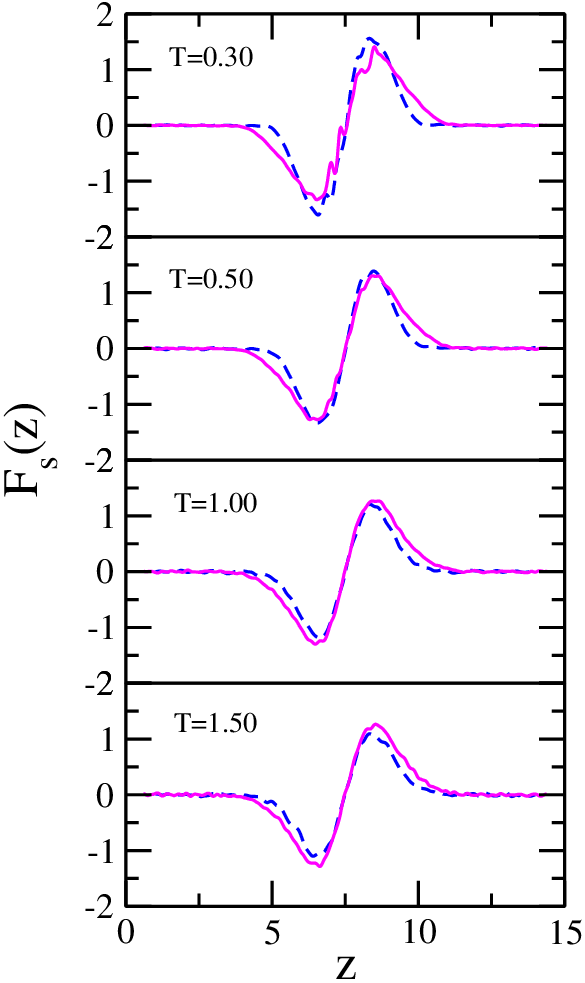}
\caption{Mean force of interaction for the small particles along the $z$ - direction when the system has only small particles(dashed lines) and when both large and small particles
present (solid lines).}
\label{fig5}
\end{figure}

\begin{figure}
\includegraphics[width=6cm, angle=-0]{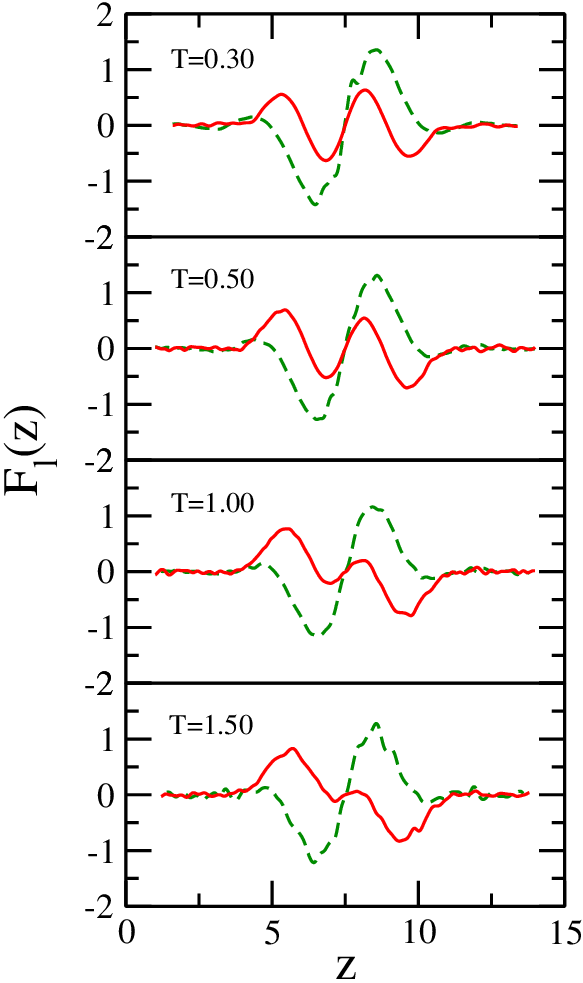}
\caption{Mean force of interaction for the large particles along the $z$ - direction when the 
system has only large particles(dashed lines) and when both large and small particles
present (solid lines).}
\label{fig6}
\end{figure}

 Waiting time distributions measure the delay times between successive hops of particles in a dynamics process.
 We have also calculated the waiting time distributions for both large and small particles from the molecular dynamics 
 trajectories by calculating the delay time between successive jumps of the particles over the barrier. 
 A log-log representation of these  waiting time distributions for both the particles
has been plotted in Figure 7. 
 As temperature decreases, the waiting time distribution for both type of particles flattens out. This is 
 expected as particles will spend more time between two potential barriers before crossing over and their
 dynamics become more and more localized as the temperature decreases.  
 However, the decay of waiting time distribution of smaller particles is much slower compared to that of larger
 particles. These distributions can be very well fitted to a sum of two exponentials.
 This essentially tells us that there are two time scales associated
 with the barrier crossing of particles. We have plotted the two relaxation times($\tau_1$ and $\tau_2$) 
 for smaller particles and the two relaxation times($\tau_3$ and $\tau_4$) 
 for larger particles against temperature in the inset of Figure 7(a) and 7(b) respectively.
 One of the relaxation times($\tau_1$ or $\tau_3$) in each case remains mostly unaffected as temperature 
 decreases; while the other ($\tau_2$ or $\tau_4$) increases as the 
 temperature decreases. $\tau_1$ (or $\tau_3$) corresponds to the recrossing occurs near the potential barriers, 
 which explains the weak dependence on temperature. 
 The larger relaxation time, $\tau_2$ (or $\tau_4$) can be attributed to the long time diffusion due to barrier
 crossing. As evident in the figure, $\tau_2$ increases rapidly with decreasing temparture while $\tau_4$ increases
 more slowly as temperature decreases. 
 This indicates increasing localization of smaller particles between the 
 repulsive barriers compared to that of larger particles and support the sub-Arrhenius diffusion obeserved in 
 the dynamics of larger particles.

 \begin{figure}
\includegraphics[width=6cm, angle=-0]{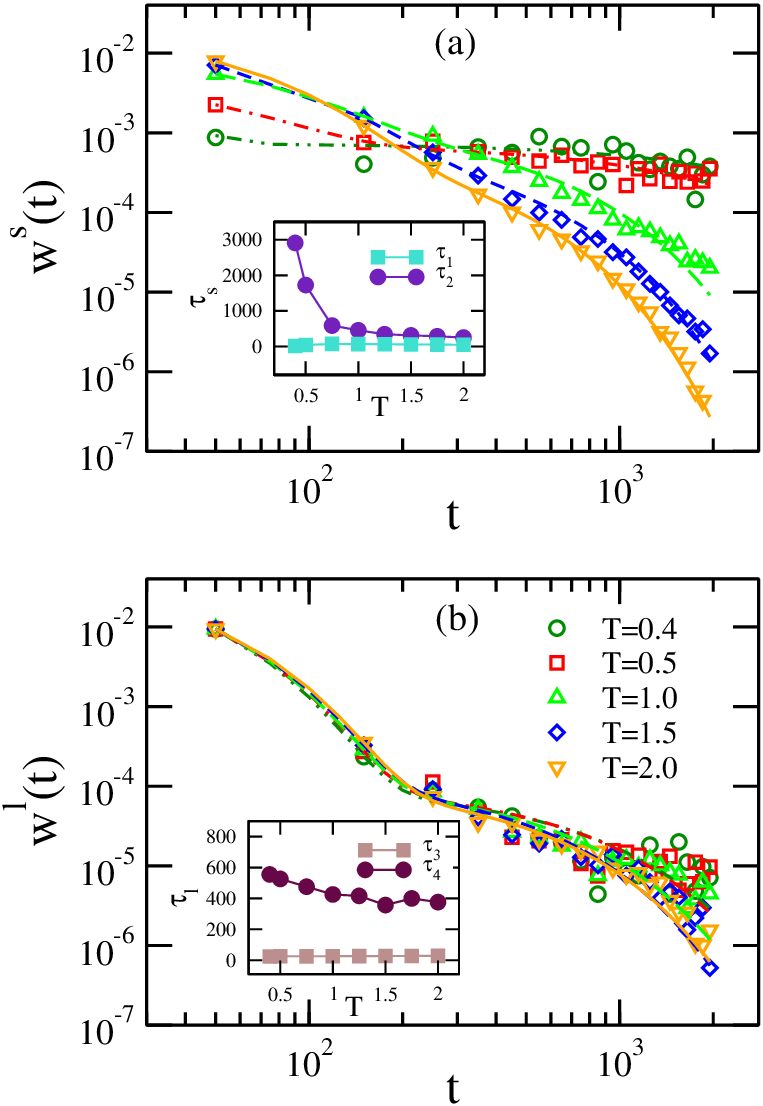}
\caption{Waiting time distribution for (a) small particles and (b) large particles for $L$=17.0. The symbols are 
the data from the simulations and the lines are exponential fits. For both type of particles
a sum of two exponentials fits the data very well. The relaxation times for each type of particles are
plotted in the inset of figure 7(a) and 7(b).}
\label{fig7}
\end{figure}
 
 It will be interesting to note that the two components of the binary mixture follow different dynamics altogether.
 The dynamics of the larger particles is mainly determined by the depletion interactions between the larger 
 particles and the potential barrier. This depletion interaction is temperature dependent and changes the 
 effective barrier accordingly. Thus, we observe a sub-Arrhenius diffusive behavior for larger particles. The 
 effective barrier of the smaller particles increases because larger particles get crowded in the region of
 external potential barrier and smaller particles gets depleted away from that region. This crowding of 
 larger particles and hence the depletion of smaller particles from the region of potential barrier does not depend 
 on temperature, especially since the volume fraction of the particles is small. So the effective potential barrier 
 for smaller particles remain unchanged with respect to temperature and they follow Arrhenius diffusion. For such 
 system slowing down of dynamics has been reported earlier\cite{dalle-ferrier}. However,
 a temperature dependent activation energy can be expected at higher density (or volume fraction) and 
 smaller particles may undergo a super-Arrhenius diffusion. Work in this direction is in progress and will be 
 reported elsewhere\cite{mustakim}. It should also be mentioned that the results are more general over a
 wide range of parameters than the set we used in these simulations\cite{mustakim}. 
 
 \section{Conclusions}
 
 We have studied the dynamics of binary colloidal mixtures, different speciens in the mixtrue differ in their sizes, subjected to 
 an external Gaussian potential by using canonical molecular dynamics simulations.
 We have shown that, contrary to the agreement in literature that sub-Arrhenius diffusion is 
 related to quantum phenomena, a classical system can show a sub-Arrhenius temperature dependence of diffusivity.
 An increase in the probability of crossing the barrier can lead to sub-Arrhenius behaviour irrespective of the 
 nature of the process involved. In this sense, the depletion interactions in the present investigation or quantum tunneling in 
 the earlier reported investigations have similar effects on the barrier crossing. Eventhough the results we 
 obtained are for a binary mixture of colloids, we believe our findings are applicable to many other systems 
 which involve barrier crossing.  Many of the biological transport process which involve more than one components, 
 differing in their dimensions, shows faster diffusivity for bigger components\cite{boda,lopez}. Similarly, anomalous 
 changes in diffusivity with respect to particle dimensions are reported for particles diffusion in porous materials
 such as zeolites\cite{turro, sarkar} and metal oxide frameworks\cite{liu, banerjee}. The pathways of 
 transport in these systems involve many bottlenecks, which can be similar to the external potential used in our 
 investigations. Therefore, our results contribute to the understanding of such processes. 
 
 The authors acknowledge the financial support from Department of Atomic Energy, India through the 12th plan project(12-R\&D-NIS-5.02-0100)

\bibliography{aps} 
\bibliographystyle{aps} 

\end{document}